\documentclass[aps,prb,twocolumn,reprint,nofootinbib,superscriptaddress,showkeys]{revtex4-1}

\usepackage{color}
\usepackage{graphicx}
\usepackage{dcolumn}
\usepackage{bm}
\usepackage{commath}
\usepackage[mathlines]{lineno}
\usepackage[english]{babel}
\usepackage[latin1]{inputenc}
\usepackage[bottom=2cm,top=2cm, left=1.5cm, right=1.5cm]{geometry}
\usepackage[unicode=true, bookmarks=true, bookmarksnumbered=false, bookmarksopen=false, breaklinks=false, pdfborder={0 0 1}, backref=false, colorlinks=false]{hyperref}
\usepackage{xcolor}

\usepackage{ulem}

\hypersetup{pdftitle={Missing spectral weight in a heavy-fermion system far above N\'eel temperature}, pdfauthor={Jingwen Li, Debankit Priyadarshi, Chia-Jung Yang, Oliver Stockert, Hilbert von L\"{o}hneysen, Shovon Pal, Manfred Fiebig, and Johann Kroha}, pdfkeywords={Strongly-correlated materials, heavy-fermions, THz time-domain spectroscopy, quantum phase transition}}

\begin{document}

\title{Missing spectral weight in a heavy-fermion system far above N\'eel temperature}

\author{Jingwen Li}
\affiliation{Department of Materials, ETH Zurich, 8093 Zurich, Switzerland}

\author{Debankit Priyadarshi}
\affiliation{Department of Materials, ETH Zurich, 8093 Zurich, Switzerland}

\author{Chia-Jung Yang}
\affiliation{Department of Materials, ETH Zurich, 8093 Zurich, Switzerland}

\author{Ulli Pohl}
\affiliation{Physikalisches Institut and Bethe Center for Theoretical Physics, University of Bonn, 53115 Bonn, Germany}

\author{Oliver Stockert}
\affiliation{Max Planck Institute for Chemical Physics of Solids, 01187 Dresden, Germany}

\author{Hilbert von L\"{o}hneysen}
\affiliation{Institut f\"{u}r Quantenmaterialien und -technologien and Physikalisches Institut, Karlsruhe Institute of Technology, 76021 Karlsruhe, Germany}

\author{Shovon Pal}
\email{shovon.pal@niser.ac.in}
\affiliation{School of Physical Sciences, National Institute of Science Education and Research, HBNI, Jatni, 752 050 Odisha, India}

\author{Manfred Fiebig}
\email{manfred.fiebig@mat.ethz.ch}
\affiliation{Department of Materials, ETH Zurich, 8093 Zurich, Switzerland}

\author{Johann Kroha}
\email{kroha@physik.uni-bonn.de}
\affiliation{Physikalisches Institut and Bethe Center for Theoretical Physics, University of Bonn, 53115 Bonn, Germany}
\affiliation{School of Physics and Astronomy, University of St.~Andrews, 
St.~Andrews, KY16 9SS, United Kingdom}

\date{\today}

\begin{abstract}
The competition between the Kondo spin-screening effect and the Ruderman-Kittel-Kasuya-Yosida (RKKY) interaction in heavy-fermion systems drives the quantum phase transition between the magnetically ordered and the heavy-Fermi-liquid ground states. Despite intensive investigations of heavy quasiparticles on the Kondo-screened side of the quantum phase transition and of their breakdown at the quantum critical point, the magnetically ordering side has not systematically been studied. Using terahertz time-domain spectroscopy, we report a suppression of the Kondo quasiparticle weight in CeCu$_{6-x}$Au$_x$ samples on the antiferromagnetic side of the quantum phase transition at temperatures as much as two orders of magnitude above the N\'{e}el temperature $T_\text{N}$. With our systematic investigations into the high-temperature, paramagnetic region on the antiferromagnetic side of the phase diagram of CeCu$_{6-x}$Au$_x$, i.e., with $x = 0.2$, 0.3 and 0.5, we show that the suppression results from a quantum frustration effect induced by the temperature-independent RKKY interaction. Hence, our results emphasize that besides critical fluctuations, the RKKY interaction may play an important role in the quantum-critical scenario.
\end{abstract}


\maketitle

\section{Introduction} 

\begin{figure*}[t!]
\centering
\includegraphics[width=0.9\linewidth]{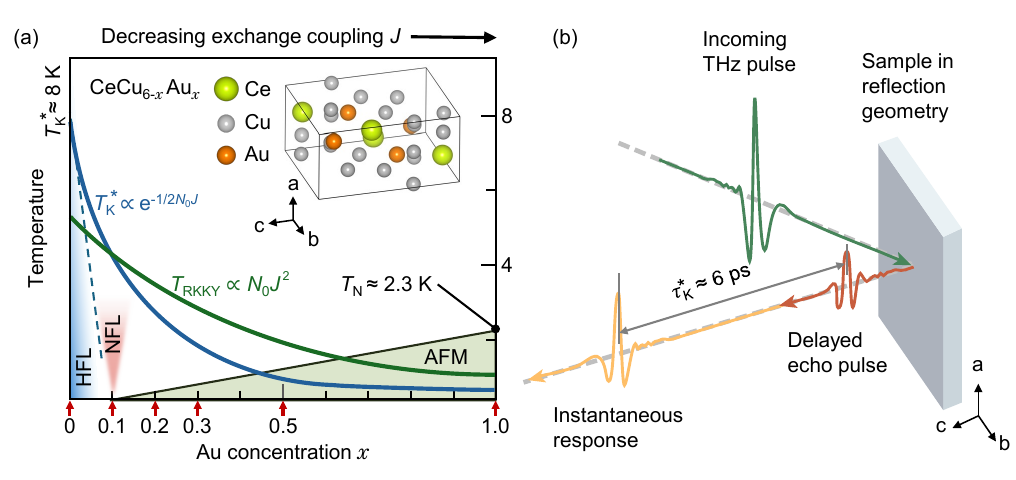}
  \caption{(a) Quantum phase transition in CeCu$_{6-x}$Au$_x$ from the HFL phase to the antiferromagnetically ordered phase, parameterized by the non-thermal control parameter $x$. While the RKKY interaction scales as $T_{\rm RKKY}\propto N_{0}J(x)^2$, the Kondo energy scales as $T_{\rm K}^*\approx D e^{-1/(2N_{0}J)}$, where $J$ decreases with the increase of $x$. The QCP occurs at the critical substitution $x_{\rm c}=0.1$ where $T_{\rm RKKY} \approx T_{\rm K}^*$. The inset shows the orthorhombic crystal structure of the CeCu$_{6-x}$Au$_x$ system. (b) Schematic of the experimental geometry showing the incident and the reflected THz pulses on the sample. Upon THz excitation, a fraction of the correlated Kondo states gets destroyed. It then takes the Kondo coherence time $\tau_{\rm K}^*$ for the reconstruction of the heavy-quasiparticle state, accompanied by the emission of a characteristic echo pulse in the THz range at a delay of $\approx\tau_{\rm K}^*$.}
\label{Figure_1}
\end{figure*}

Heavy-fermion materials are prototypes of strongly interacting systems that contain a matrix of localized magnetic moments in $4f$ orbitals, immersed in a sea of mobile conduction electrons~\cite{ColemanHF,Loehneysen2007,HewsonHF}. The hybridization of localized $4f$ electrons with conduction electrons generates (i) the Kondo spin-exchange coupling $J$ between local and itinerant moments~\cite{Anderson1961}, and (ii) in the second order of $J$, a long-range correlation between distant local moments, the Ruderman-Kittel-Kasuya-Yosida (RKKY) interaction~\cite{Ruderman1954,Kasuya1956,Yosida1957}. The local moments are screened by the conduction spins due to the Kondo effect~\cite{HewsonHF,Kondo1964}. In contrast to charge screening, spin screening is a quantum effect which leads to a many-body entangled, singlet ground state below the binding energy. The latter is given roughly by~\cite{HewsonHF} the lattice Kondo temperature $T_{\rm K}^*\approx De^{-1/(2N_0J)}$, where $D$ is the half bandwidth, and $N_0$ is the conduction density of states at the Fermi level $E_{\rm F}$. The excitations of this state are complex quantum superpositions of localized and itinerant states, the heavy, fermionic quasiparticles which form a narrow band near $E_{\rm F}$~\cite{HewsonHF,Li2023}. On the other hand, the RKKY interaction, which scales as $T_{{\rm RKKY}}\approx N_0J^2$, tends to order the magnetic moments and, thus, counteracts the Kondo spin screening. Therefore, Doniach~\cite{Doniach1977} put forward a schematic heavy-fermion phase diagram which consists of a spin-screened, heavy-Fermi-liquid (HFL) phase realized at low temperatures $T\lesssim T_{\rm K}^*$ when the Kondo scale dominates over the RKKY energy, $T_{\rm K}^* > T_{\rm RKKY}$, and a magnetically ordered phase for $T_{\rm RKKY}>T_{\rm K}^*$. In the heavy-fermion compound CeCu$_{6-x}$Au$_{x}$, for example, substituting Cu by Au reduces the spin-exchange coupling $J$ and, thus, leads to the phase diagram shown in Fig.~\ref{Figure_1}(a); see below for more details. The HFL and the antiferromagnetic (AFM) phases are separated by a quantum critical point (QCP) at $x=x_{\rm c}=0.1$ where $T_{\rm K}^*$ and $T_{\rm RKKY}$ balance each other.

The behavior of the heavy, fermionic quasiparticles near the quantum phase transition (QPT) has been a subject of intense discussion, since their breakdown can give way to the formation of novel quantum states of matter beyond the Fermi-liquid paradigm. Quantum-critical scenarios driven by different types of critical fluctuations have been proposed for the quantum phase transition across the quantum-critical point (QCP) in heavy-fermion compounds. These include local quantum criticality~\cite{Si2001,Coleman2001,Si2003}, Fermi surface fluctuations~\cite{Senthil2004}, or long-range antiferromagnetic fluctuations~\cite{Woelfle2011,Abrahams2014,Kambe1996,Knafo2009,Woelfle2016}. In the spin-density-wave scenario, the HFL is formed below $T_{\rm K}^*$ and undergoes a spin-density-wave instability at a critical spin coupling or Fermi-surface nesting. In this case, the QPT is driven by the critical fluctuations of the bosonic order parameter (staggered magnetization) only, and the fermionic quasiparticles remain intact across the transition~\cite{Hertz1976,Millis1993,Moriya1985}. In the local quantum-critical scenario~\cite{Si2001,Coleman2001,Si2003}, order-parameter fluctuations couple to the local Kondo spins and destroy the heavy fermionic quasiparticles near the QPT. On the magnetically ordering side of the QPT, the importance of magnon fluctuations, including a renormalization of the Kondo scale, has been analyzed by scaling theory~\cite{Irkhin1997,Irkhin2000}. Furthermore, it was shown generally~\cite{Nejati2017} and supported by numerical renormalization-group calculations for two-impurity systems with conduction-electron-induced RKKY coupling~\cite{Wojcik2023a,Wojcik2023b} that complete spin screening ceases to exist beyond a critical RKKY coupling strength. This is caused by a quantum frustration effect, which subsequently gives way to the magnetic ordering of the incompletely screened spins.  

Terahertz time-domain spectroscopy (THz-TDS) provides a unique, background-free means to investigate the spectral weight of heavy quasiparticles~\cite{Wetli2018,Pal2019,Yang2023,Yang2023np}. Specifically, the emission of a characteristic, time-delayed echo pulse after an irradiated THz pulse separates the correlation-induced Kondo spectral weight from other contributions. In the present work, we use this THz-TDS approach to systematically scrutinize the behavior of the quasiparticle weight in the prototypical heavy-fermion compound CeCu$_{6-x}$Au$_x$~\cite{Loehneysen1994}. We cover the complete range of Au concentrations, $0.0\le x \le 1.0$, in particular, covering the region of supercritical Au concentration ($x > x_{\rm c}$) and above the AFM ordering temperature $T_{\rm N}$. 

On the HFL side of the QPT, where the Kondo scale is set by $T_{\rm K}^*\approx 8$\,K~\cite{Loehneysen2007,Wetli2018}, the heavy-quasiparticle weight starts to build up already at temperatures as high as $\sim 130$\,K due to the logarithmic onset of Kondo correlations~\cite{Wetli2018}. Therefore, one might expect the same behavior on the AFM side far above $T_{\rm N}$, where AFM critical fluctuations are absent. Surprisingly, however, our experiments show that in this strongly substituted regime, the Kondo weight is systematically suppressed, even far above $T_{\rm N}$. As we will discuss, this is a clear indication for the importance of the temperature-independent RKKY interaction for the QPT in this system. It has been shown earlier that any disorder resulting from the substitution of Cu by Au, plays no role on long-range magnetic ordering in the CeCu$_{6-x}$Au$_x$ system~\cite{Schroeder1996,Loehneysen1996,Loehneysen1998,Schroeder2000,Loehneysen2000}. As a consequence, such disorders would not hinder either the screening effects or the frustration effect that stems from the temperature-independent RKKY interactions in our systems.

\begin{figure*}[t!]
\centering
\includegraphics[width=0.95\linewidth]{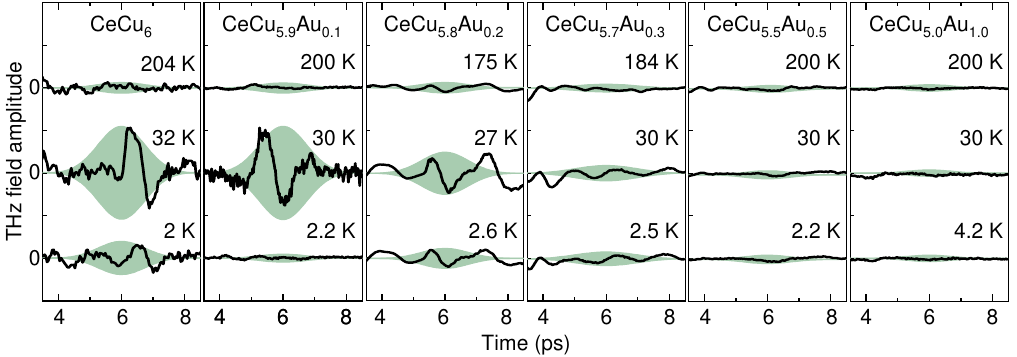}
  \caption{Normalized THz time traces in reflection from the CeCu$_{6-x}$Au$_x$ samples in three different temperature ranges. The green-shaded area depicts the envelope of the time-delayed THz reflex, containing the coherent contributions from the heavy quasiparticles. The envelope is obtained from the solution of the nonlinear rate equation as discussed in Ref.~[\onlinecite{Wetli2018}]. Note that a temperature-independent background signal is present at all temperatures. It has been determined close to room temperature where the Kondo weight is undetectably small, and subtracted from the signal at all temperatures~\cite{Wetli2018}.}
\label{Figure_2}
\end{figure*} 

\section{Samples and experimental method}

Single crystals of CeCu$_{6-x}$Au$_x$ were grown by the Czochralski method~\cite{Loehneysen1994,Loehneysen1993,Schlager1993,Neubert1997} and are freshly polished using colloidal silica before the THz-TDS measurements~\cite{Wetli2018,Pal2019}. The surface roughness during sample preparation is within the sub-micrometer range, significantly below the wavelength of THz radiation. The concentrations of Au used in this work are $x = 0.0,~0.1,~0.2,~0.3,~0.5,$ and $1.0$, marked by the red arrows in Fig.~\ref{Figure_1}(a). The samples are mounted in a Janis SVT-400 helium reservoir cryostat with a controlled temperature environment ranging from 300\,K down to 2\,K. 

Substitution of Cu by the larger Au atoms expands the host lattice while maintaining the orthorhombic structure, see inset of Fig.~\ref{Figure_1}(a). This reduces the Ce~$4f$ conduction-electron hybridization and, hence, the exchange coupling $J$. Considering the parametric dependence of  $T_{\rm K}^*$ and $T_{\rm RKKY}$ on $J$ and following Doniach's argument~\cite{Doniach1977}, below the critical Au concentration of $x<x_{\rm c}=0.1$, a spin-screened HFL phase is stabilized below the Kondo-scale temperature $T_{\rm K}^*$~\cite{Loehneysen2007,Amato1987}. For $x>x_c$, an incommensurate AFM phase is realized~\cite{Loehneysen1998}. For $x=1.0$, the N\'eel temperature is $T_{\rm N}\approx 2.3$\,K with a linear decrease of $T_{\rm N}$ as a function of $x$ as the QCP at $x_{\rm c}=0.1$ is approached~\cite{Loehneysen2007,Pietrus1995}.

In the experiment, single-cycle, linearly polarized THz pulses are generated via optical rectification in a 0.5-mm-thick (110)-cut ZnTe generation crystal, using up to 90\% of an amplified Ti:sapphire laser output (wavelength of 800\,nm, pulse duration of 120\,fs, repetition rate of 1\,kHz, pulse energy of 1.5\,mJ). The THz light pulses are then guided onto the sample using off-axis parabolic mirrors. As the samples are metallic, we collect the signal in reflection geometry, as shown in Fig.~\ref{Figure_1}(b). We measure both the time-dependent amplitude and phase of the reflected THz light by free-space electro-optic sampling, using the residual 10\% of the laser output as the sampling light. The THz and the sampling beams are collinearly focused onto a 0.5-mm-thick, (110)-cut ZnTe detection crystal. The THz-light-induced ellipticity of the sampling beam is then measured using a quarter-wave plate, a Wollaston prism, and a balanced photodiode. In order to increase the accessible delay time between the THz pump and the optical probe pulses, Fabry-P\'erot resonances from the faces of the detection crystal are suppressed by an additional 2-mm-thick, THz-inactive (100)-cut ZnTe single crystal that is optically bonded to the back of the detection crystal. All measurements are performed in an inert N$_{2}$ atmosphere.

\section{Results and Discussion}

When our heavy-fermion material interacts with the THz radiation, a fraction of the correlated Kondo states gets destroyed~\cite{Wetli2018,Pal2019,Yang2023,Yang2023np}. In other words, the excitation leads to an electronic interband transition from the heavy-fermion band into the light part of the conduction band~\cite{Yang2020}, breaking the Kondo singlet and thereby deleting the associated spectral weight of the heavy band. The reconstruction of the heavy-quasiparticle state happens on the scale of the Kondo coherence time $\tau_{\rm K}^* = 2\pi \hbar /k_{\rm B} T_{\rm K}^*$, when the relaxing electrons emit a characteristic, echo-like pulse at a delay time $\tau_{\rm K}^*$. The echo response is background-free and bears distinct information on the Kondo correlation dynamics~\cite{Wetli2018,Yang2023np}. It is a fingerprint of the electronic coherence that is intrinsic to the heavy bands~\cite{Yang2024} and is separated in the time-domain from the instantaneous response which arises from intraband excitations, see the schematic illustration in Fig.\,\ref{Figure_1}(b). The echo-pulse intensity and delay time are measures of the quasiparticle weight and their intrinsic coherence time $\tau_{\rm K}^*$, respectively. A quantitative comparison between the CeCu$_{6-x}$Au$_x$ samples with different Au concentrations is then obtained by normalizing the THz signals with respect to the area of the samples.

\begin{figure*}[t!]
\centering
\includegraphics[width=0.95\linewidth]{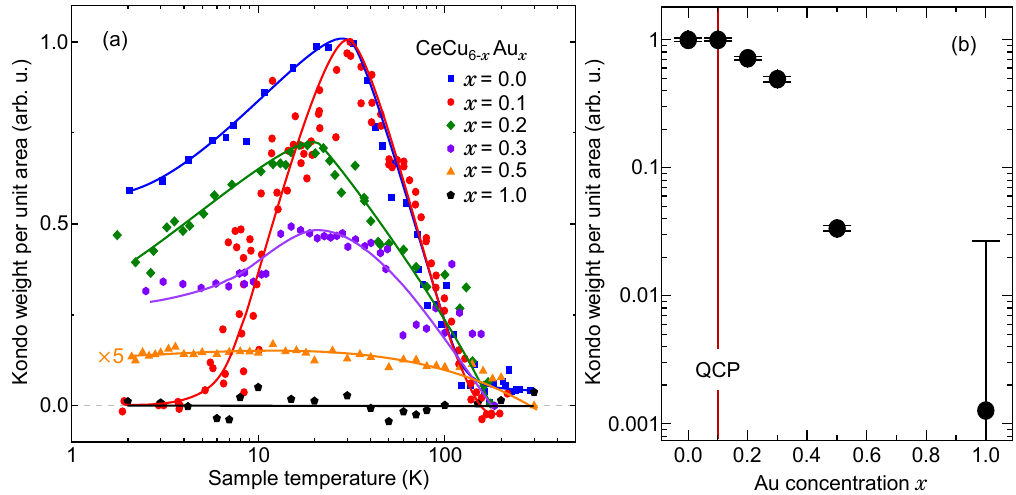}
  \caption{(a) Temperature dependence of the Kondo weight per unit area of our CeCu$_{6-x}$Au$_x$ samples. Published data~\cite{Wetli2018} ($x=0.0, 0.1, 1.0$) are shown along with our new data ($x=0.2, 0.3, 0.5$) for better accentuation of the different responses on the low- and high-substitution sides of the QCP. The weight is derived from the integrated intensity of the reflected THz echo pulse, see Fig.~\ref{Figure_2}. The solid lines are guides to the eye. (b) The Kondo spectral weight per unit area of CeCu$_{6-x}$Au$_x$ as a function of Au concentration $x$ at the temperature associated with the fitted maximum of the Kondo weight in (a) (except for $x=1$ where we average the entire data set).}
\label{Figure_3}
\end{figure*}

To navigate through the phase diagram, we investigate the CeCu$_{6-x}$Au$_x$ sample with $x=0$, exhibiting an HFL ground state, the quantum-critical sample with $x=0.1$, and samples with $x=0.2$, 0.3, 0.5, and 1.0, featuring an AFM ground state~\cite{Loehneysen2007,Amato1987,Loehneysen1998}. We deliberately show our measured data for $x=0.2,~0.3,~0.5$ along with published data at  $x=0.0,~0.1,~1.0$~\cite{Wetli2018} because only the direct comparison of the two sets reveals the stark difference in the response on the low- and high-substitution sides of the QCP with a previously unrecognized behavior on the latter side.

Fig.~\ref{Figure_2} shows the traces of the THz transients reflected from our samples for three representative temperature ranges, each at a fixed time window containing the echo pulse. These data are normalized to the total reflected power (including the instantaneous reflex) and to the sample area~\cite{Wetli2018}. The quasiparticle spectral weight for all the samples is calculated in the following way: The Kondo spectral weights at different temperatures are calculated by integrating the squared-electric field of the normalized time traces over the interval from 3.5\,ps to 8.5\,ps, where the echo pulse appears. To remove the incoherent temperature-independent background in the Kondo spectral weight~\cite{Wetli2018}, we subtract the THz time-domain response at the highest temperature of each measurement. The analysis is discussed in more detail in Ref.~[\onlinecite{Wetli2018}].

For the heavy-fermion (CeCu$_{6}$) and the quantum-critical (CeCu$_{5.9}$Au$_{0.1}$) compounds, the spectral weight builds up from around 130\,K and increases logarithmically down to 30\,K as the manifestation of the Kondo effect. Upon further reducing the temperature, the spectral weight of the quantum-critical sample drops continuously and reaches an undetectably small value for $T < 5$\,K, indicating quasiparticle disintegration near the QCP~\cite{Wetli2018}. In CeCu$_{6}$, the spectral weight reaches the same maximum as for the quantum-critical sample, yet it settles to a finite value toward the lowest temperatures, roughly 40\% beneath its maximum. This finite, but decreased spectral weight reflects the HFL ground state in proximity to the QCP.

Beyond the QCP, that is, for the CeCu$_{6-x}$Au$_x$ samples with $x > 0.1$, the temperature dependence of the data recorded for the present work exhibits a strikingly different and quite surprising behavior in comparison to the data at $x < 0.1$. We observe a pronounced decrease in the maximum Kondo weight with increasing $x$. While in CeCu$_{5.8}$Au$_{0.2}$ as well as in CeCu$_{5.7}$Au$_{0.3}$ the Kondo spectral weight shows a similar onset near 130\,K, its build-up weakens as $x$ increases, and it reaches a maximum lower than for the HFL sample ($x=0$), before it settles to a reduced value at the lowest temperatures; see also Ref.~[\onlinecite{Klein2008}]. At $x=0.5$, the Kondo weight is still non-zero, yet barely detectable, and it has vanished entirely at $x = 1.0$. This $x$ dependence of the Kondo weight is shown in Fig.~\ref{Figure_3}(b) for the respective fitted maximum of the temperature-dependent Kondo-weight data in Fig.~\ref{Figure_3}(a) (except for $x=1$ where we average the entire data set). While the decrease in maximum weight with $x$ is moderate for $x = 0.2$ and $x = 0.3$, a steep decline occurs beyond $x = 0.3$. We suspect that this reflects the $x$-dependent exponential decrease of the Kondo energy with respect to the RKKY interaction, $T_{\rm K}^*/T_{\rm RKKY}$, as sketched in Fig.~\ref{Figure_1}, but further samples and analysis would be required to support this assumption.

\begin{figure}[t!]
\centering
\includegraphics[width = \columnwidth]{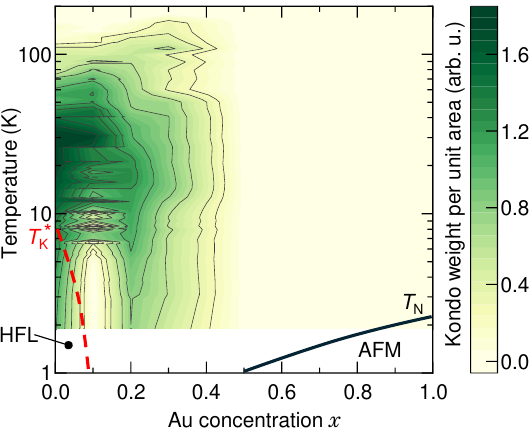}
  \caption{Phase diagram of CeCu$_{6-x}$Au$_x$ obtained by tuning the Au concentration. The color code represents the heavy-quasiparticle spectral weight. Near the QCP, the quasiparticle weight decreases as a result of critical fluctuations. In contrast, the missing heavy quasiparticle spectral weight beyond $x\approx 0.4$ is caused by the RKKY-induced frustration of Kondo-singlet formation.}
\label{Figure_4}
\end{figure}

Fig.~\ref{Figure_4} shows the phase diagram of CeCu$_{6-x}$Au$_x$ where the Kondo spectral weight, represented through by the color code, has been interpolated from the samples with $x=0,~0.1,~0.2,~0.3,~0.5,~{\rm and}~1.0$ available for this study. The figure clearly demonstrates that the Kondo destruction at $x\gtrsim 0.4$ occurs in the high-temperature region of the phase diagram where critical AFM order-parameter fluctuations are absent. The destruction must, therefore, be caused by the RKKY interaction alone, which is active at all temperatures. Such RKKY-induced breakdown has been predicted in Refs.~[\onlinecite{Nejati2017}] and [\onlinecite{Wojcik2023a}] and is due to a quantum frustration effect between Kondo spin screening and inter-impurity RKKY interaction. Specifically, when coupled to the spin of a distant Kondo ion, a local impurity spin cannot freely fluctuate to form a singlet with the surrounding conduction electrons. Similar Kondo destruction has been found in scanning-tunneling spectroscopy experiments on CoCu$_n$Co clusters ($n \geq 3$) on Cu(111)~\cite{Neel2011} and on a continuously tunable two-impurity Kondo system of Co adatoms at the tip and on Au(111)~\cite{Bork2011}. We observe this effect here in a bulk Kondo-lattice system. We would also like to stress that the Kondo breakdown region around $x = 0.4$ is well separated from the quantum-critical region around $x_{\rm c}=0.1$. It indicates that the effect of RKKY-related quantum frustration and the effect of critical fluctuations emerge independently. Hence, the QPT is most probably induced by the cooperation of RKKY-induced frustration of spin screening and critical fluctuations of the AFM order parameter~\cite{Si2001,Coleman2001,Si2003}.

\section{Conclusion}

In conclusion, we have investigated the behavior of the heavy-quasiparticle weight across the quantum phase transition induced in the CeCu$_{6-x}$Au$_x$ system by tuning the Au concentration. While a logarithmic build-up of spectral weight is observed when the Kondo effect dominates over the RKKY interaction ($x \le 0.1$), in the CeCu$_{6-x}$Au$_x$ compounds with $x > 0.1$ the heavy quasiparticle spectral weight is significantly suppressed and essentially vanishes for $x > 0.5$. Most unusually, this behavior occurs at temperatures much higher than the magnetic ordering temperature, that is, in the paramagnetic region of the antiferromagnetic side of the phase diagram. Our results point to the importance of the frustration effect of the RKKY interaction on the formation of heavy-fermion singlets. The separation of the RKKY-induced breakdown from the quantum phase transition at $x_{\rm c}=0.1$ in the phase diagram suggests that the former emerges independent of the latter, implying that the QPT is the result of cooperation of both the RKKY-induced frustration of the spin screening and the critical fluctuations of the AFM order parameter.

\section{Acknowledgement} 

J.L., D.P., C.-J.Y., and M.F. acknowledge the financial support from the Swiss National Science Foundation through Projects No. 200021\_178825 and No.\ 200021\_219807. S.P. acknowledges the support from DAE through the project Basic Research in Physical and Multidisciplinary Sciences via RIN4001. S.P. also acknowledges the startup support from DAE through NISER and SERB through SERB-SRG via Project No.\ SRG/2022/000290. J.K. acknowledges the support of the Deutsche Forschungsgemeinschaft via TRR 185 (277625399) OSCAR, project C4, and the Cluster of Excellence ML4Q (90534769). 

S.P., M.F., and J.K.\ contributed equally to this work.

\end{document}